\documentclass[12p]{article}
\usepackage{latexsym}
\usepackage{color}
\usepackage{amsmath}
\usepackage{epsfig}
\usepackage{hyperref}
 \usepackage{dcolumn}
\newcommand{\ortala}[1]{\begin{center}#1\end{center}}

\newcommand{\sandd}[1]{\left\langle #1\right\rangle}

\newcommand{\intego}[3]{{{\underset{#1
}{\overset{#2}{\displaystyle\oint}}}#3}}
\newcommand{\summ}[3]{{{\underset{#1 }{\overset{#2}{\displaystyle\sum}}}#3}}

\newcommand{\re}[1]{(\ref{#1})}

\newcommand{\eq}[2]{\begin{equation}\label{#1}  #2\end{equation}}
\newcommand{\tur}[2]{\frac{d#1}{d#2}}

\newcommand{\paran}[1]{\left(#1\right)}

\newcommand{\sch}[1]{Schrodinger}

\newcommand{\komb}[2]{\paran{\begin{array}{c} #1 \\ #2 \end{array}}}
\newcommand{\sek}[1]{Fig. \ref{#1}}

\linethickness{0.6pt}

\begin{document}

\ortala{\textbf{ Dynamical response of the Ising model to the amplitude 
modulated time dependent magnetic field }}

\ortala{\textbf{\"Umit Ak\i nc\i \footnote{umit.akinci@deu.edu.tr}}}

\ortala{\textit{Department of Physics, Dokuz Eyl\"ul University,
TR-35160 Izmir, Turkey}}

\section*{Abstract}

The dynamical Ising model under the effect of the amplitude modulated time 
dependent periodic magnetic field has been solved
by using EFT with Glauber type of stochastic process. Several cases with 
amplitude modulation have been investigated. It has been shown 
that, amplitude modulation could display dynamical phase transition on the 
magnetic system.

Keywords: \textbf{Dynamic Ising model;
amplitude modulated field, dynamical phase transition}

\section{Introduction}\label{introduction}

Dynamical response of the spin system to the time dependent magnetic field has 
attract attention both theoretically
and experimentally. Relation  between the relaxation time of the spin system 
and 
period of the driving periodic 
external magnetic field determines the dynamic phase of the system.  Dynamic 
phase transition (DPT) in these systems 
come from the competition  between these two time scales  \cite{ref1}. 

Theoretically DPT in spin systems 
first observed within the mean field approximation (MFA) \cite{ref2} for the 
spin-$1/2$ Ising model. 
After that time, DPT in the spin-$1/2$ Ising model has been widely studied 
within the
several techniques such as MFA \cite{ref3}, Monte Carlo 
simulation (MC) \cite{ref4},
effective field theory (EFT) \cite{ref5}.  Dynamical character was mostly included 
within the 
Glauber-type stochastic process \cite{ref6} to these techniques.   

Experimentally, DPT can be observed in several systems such as,
ultrathin Co films \cite{ref7},  Fe/Au(001)
films \cite{ref8}, epitaxial Fe/GaAs(001) thin
films \cite{ref9},  fcc Co(001), and fcc NiFe/Cu/Co(001) layers \cite{ref10} 
and 
Fe/InAs(001) ultrathin films \cite{ref11}.

Theoretical efforts continuing with different type of spin systems. Higher spin 
Ising models  such as spin-1 Blume-Capel model \cite{ref12} and spin-$3/2$ 
Ising 
model \cite{ref13} have been studied as well as spin systems  with quenched 
disorder; random fields \cite{ref14}, random crystal fields \cite{ref15}, bond 
dilution \cite{ref16} and site dilution \cite{ref17}. Besides dynamical 
properties of the Ising model on different geometries have been studied. 
Nanotube \cite{ref18}, nanowire \cite{ref19}, thin film \cite{ref20} and Bethe 
lattice \cite{ref21} geometries are among them.  Although not as common as 
Ising 
model, dynamical properties of the other models do exist. Heisenberg model 
\cite{ref22} and Ashkin-Teller \cite{ref23} model have been studied by means of 
MC and EFT with Glauber type of stochastic process, respectively.

On the other hand, special attention was paid on the different types of external 
time dependent magnetic fields. These different types of driving fields are 
promising 
diverse dynamical phase transition and hysteresis properties. Ising 
ferromagnets 
under the effect of the pulsed \cite{ref24,ref25} and randomly varied external 
time dependent fields \cite{ref26} have been studied within the MC. For the 
system with pulsed field, it has been shown that, ratio of time width of the 
response (i.e. time dependent magnetization) of the system to the time width of 
the pulse field diverges at the critical temperature of the static system 
\cite{ref25}. Besides, ratio between the response magnetization peak height and 
the pulse height gives peak near this temperature \cite{ref25}.  For the 
randomly varied field it has been shown that, if the interval of the 
distribution of which random field chosen large enough, dynamically disordered 
phase can be created \cite{ref26}. These results have been obtained within the 
MC simulation and by solving MFA equations created with Glauber type of 
stochastic process. Randomly varied (fast switching) external field problem was 
also 
studied  with slightly different approach \cite{ref27,ref28}.  Besides, effect 
of the polarized electromagnetic wave \cite{ref29} and field which have 
gradient 
\cite{ref30} inspected by the same methods. Other examples of this category 
are, 
 standing wave \cite{ref31} and propagating wave type of fields  
\cite{ref32,ref33}.

Indeed dynamical spin system under the different type of fields deserves more 
attention, due to the rich dynamical behavior and possible experimental 
applications.   For this aim this work is devoted to the exploration of the 
dynamical 
properties of the Ising model which drives with external periodic magnetic 
field 
with amplitude modulation. Amplitude modulation is mostly used in communication 
systems and it is not a difficult task to obtain modulated signal. The formulation 
is EFT 
with Glauber type of stochastic process \cite{ref6}. The paper is 
organized
as follows: In Sec. \ref{formulation} we briefly present the
model and  formulation. The results and discussions are
presented in Sec. \ref{results}, and finally Sec. \ref{conclusion} contains our 
conclusions.

\section{Model and Formulation}\label{formulation}

Let the periodic magnetic field with  frequency  $\omega$ $H_0\cos{\omega t}$ acts 
on the Ising spin system. 
Amplitude modulated magnetic field can be defined by
\eq{denk1}{
H(t)=\left[H_0+r(t)\right]\cos{\paran{\omega t}},
} where $r(t)$ is the time dependent field. Here, the term $H_0+r(t)$ is called as 
modulating field, while $H(t)$ given by \re{denk1} as modulated field. Choosing 
modulating field as sinusoidal form with 
frequency $\omega_m$ is named as tone modulation in communication 
literature\cite{refelek}. 
Then the total magnetic field is given by
\eq{denk2}{
H(t)=\left[H_0+H_1\cos{\paran{\omega_m t}}\right]\cos{\paran{\omega t}}.
} The (Fourier) spectrum of this field consists of frequencies $\pm \omega, 
\omega \pm \omega_m , -\omega \pm \omega_m $.


Although, the period of the magnetic field is not necessary for finding the 
time 
series of the magnetization, in order to calculate 
the dynamical order parameter we need the period of the magnetic field. In 
general, let  $f(x)$ and $g(x)$ be the 
periodic functions with periods  $p$ and $q$, respectively. The period of 
product (or sum) of these two functions is defined by 
$r=ap=bq$ (the smallest common multiple), where $a$ and $b$ are the positive 
integers. Note that $r$ need not to be the smallest
period, but this could not change the value of the dynamical order parameter.


The Hamiltonian of the dynamical Ising model is given by
\eq{denk3}{\mathcal{H}=-J\summ{<i,j>}{}{S_i S_j}
-H(t)\summ{i}{}{S_i},}
where the first summation is over the nearest neighbors of the lattice, 
while the
second one is over all the lattice sites. Here,  $S_i$ is the $z$ 
component of the spin variables at a site $i$, $J$ is the exchange 
interaction and $H(t)$ is the external magnetic field which is given by Eq. 
\re{denk2}.



Glauber-type stochastic process \cite{ref6} can be used for
investigating dynamic properties of the considered system. In general, in the
Glauber type of stochastic process (as done in Ref. \cite{ref2} for the 
mean field approximation) the thermal average (denoted with $\sandd{}$) of a 
spin variable 
$S_i$, which can take values $\pm 1$  
can be given as
\eq{denk4}{
\theta \tur{\sandd{S_i}}{t}=-\sandd{S_i}+\sandd{\frac{Tr_i S_i 
\exp{\paran{-\beta \mathcal{H}_i}}}{Tr_i \exp{\paran{-\beta \mathcal{H}_i}}}}.
}
 Here, $\theta$ is the  single spin flip rate per
unit time, $\beta=1/(k_BT)$, $k_B$ and $T$ denote the Boltzmann constant and
temperature, respectively. $Tr_i$ stands for the trace operation over the site 
$i$. Also $\mathcal{H}_i$ denotes the part of the Hamiltonian of the system 
related to the site $i$, which is given by,

\eq{denk5}{
\mathcal{H}_i=-S_i\paran{J\summ{j=1}{z}{ S_j}+
H(t)}=-S_i\paran{h_i+H(t)}
} where $z$ is the number of nearest neighbor sites  and $h_i$ is
the local field that represents the nearest neighbor interactions of a site 
$i$. 
Inserting  Eq. \re{denk5}
into Eq. \re{denk4} yields,

\eq{denk6}{
\theta \tur{\sandd{S_i}}{t}=-\sandd{S_i}+\sandd{\frac{Tr_i S_i 
\exp{\paran{\beta S_i\paran{h_i+H(t)}}}}{Tr_i \exp{\paran{\beta 
S_i\paran{h_i+H(t)}}}}}.
} Performing trace operation over degrees of freedom $S_i=\pm 1$ 
\eq{denk7}{
\theta 
\tur{\sandd{S_i}}{t}=-\sandd{S_i}+\sandd{\tanh\paran{\beta\paran{h_i+H(t))}}}
} can be obtained. Since the effect of the exponential differential operator on 
any function $F(x)$ is defined by
\eq{denk8}{
\exp{}\paran{a\nabla}F(x)=F(x+a),
} for any constant $a$ and $\nabla=d/dx$, Eq. \re{denk7} can be written by 
within differential operator technique \cite{ref34} as

\eq{denk9}{
\theta 
\tur{\sandd{S_i}}{t}=-\sandd{S_i}+\sandd{\exp{}\paran{h_i\nabla}}f(x+H(t))|_{x=0
},
}
where
\eq{denk10}{
f(x)=\tanh\paran{\beta x}.
} The last term in Eq. \re{denk9} can be evaluated within the decoupling 
approximation and this will yield 


\eq{denk11}{
\theta 
\tur{m}{t}=-m+\summ{n=0}{z}{}A_nm^n,
} where $m=\sandd{S_i}$ and

\eq{denk12}{
A_n=\frac{1}{2^z}\komb{z}{n}\summ{r=1}{z-n}{}\summ{s=0}{n}{}\komb{z-n}{r}\komb{n
}{s}\paran{-1}^sf\left[\paran{z-2r-2s}J\right]
}
This first order differential equation 
can be solved by standard methods such as the Runge-Kutta method 
\cite{ref35}.

Dynamical order parameter of the system can be defined as integration of the 
time dependent magnetization over one period ($P$) of the 
magnetic field,
\eq{denk13}{Q=\frac{1}{P}\intego{}{}{m\paran{t}dt}.} Since the determination of 
the period of the modulated magnetic field 
(defined in Eq. \re{denk2}) will yield multiples of the period ($nP$, where $n$ 
is integer), integral in Eq. \re{denk13} will be 
taken over the time $nP$. In this case $P$ will be replaced by $nP$ in Eq. 
\re{denk13}.

\section{Results and Discussion}\label{results}

The dimensionless quantities given by 
\eq{denk14}{\tau=\frac{k_BT}{J}, 
h_0=\frac{H_0}{J},h_1=\frac{H_1}{J},h(t)=\frac{H(t)}{J}.
} will be used throughout the work. Our investigation will be focused on square  
($z=4$) lattice. 
We set $\theta=1$ throughout our numerical calculations.

\subsection{Unmodulated case}\label{results1}

\begin{figure}[h]\begin{center}
\epsfig{file=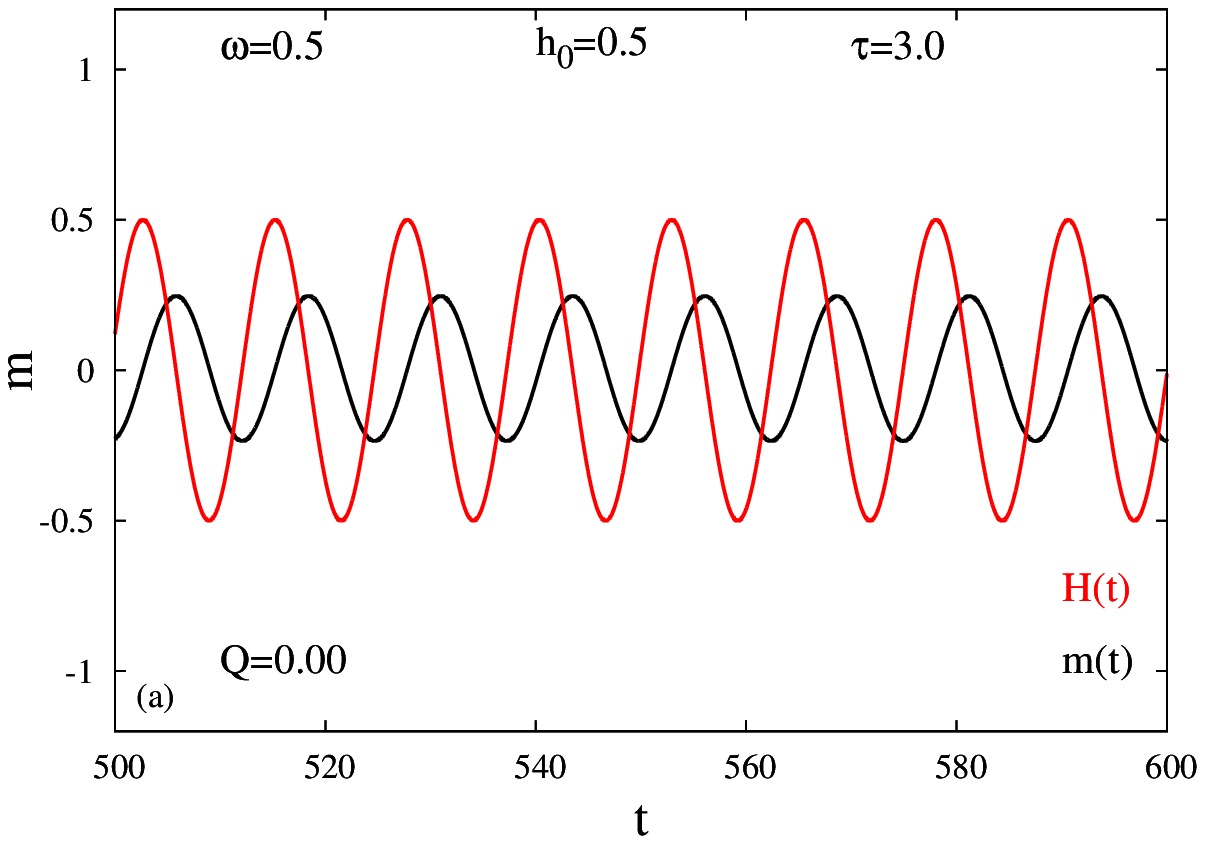, width=7.4cm}
\epsfig{file=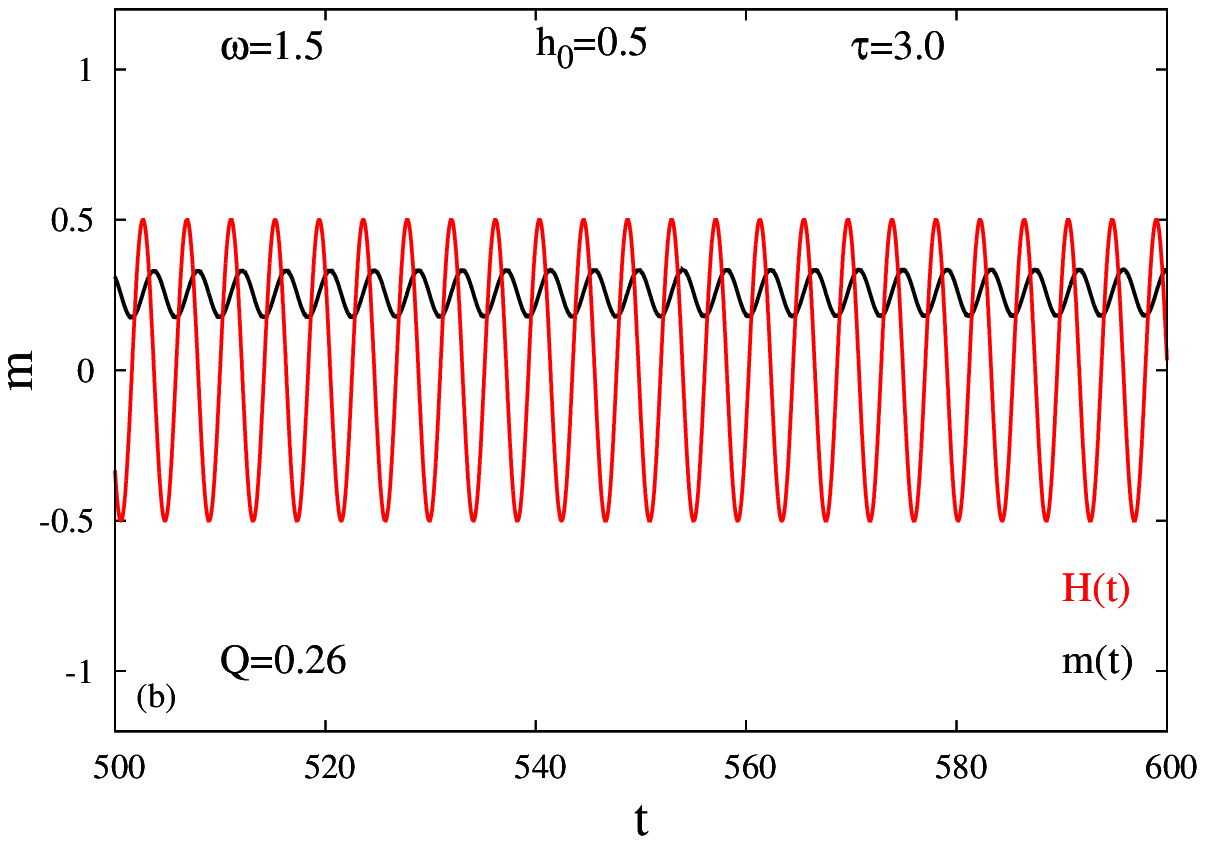, width=7.4cm}
\epsfig{file=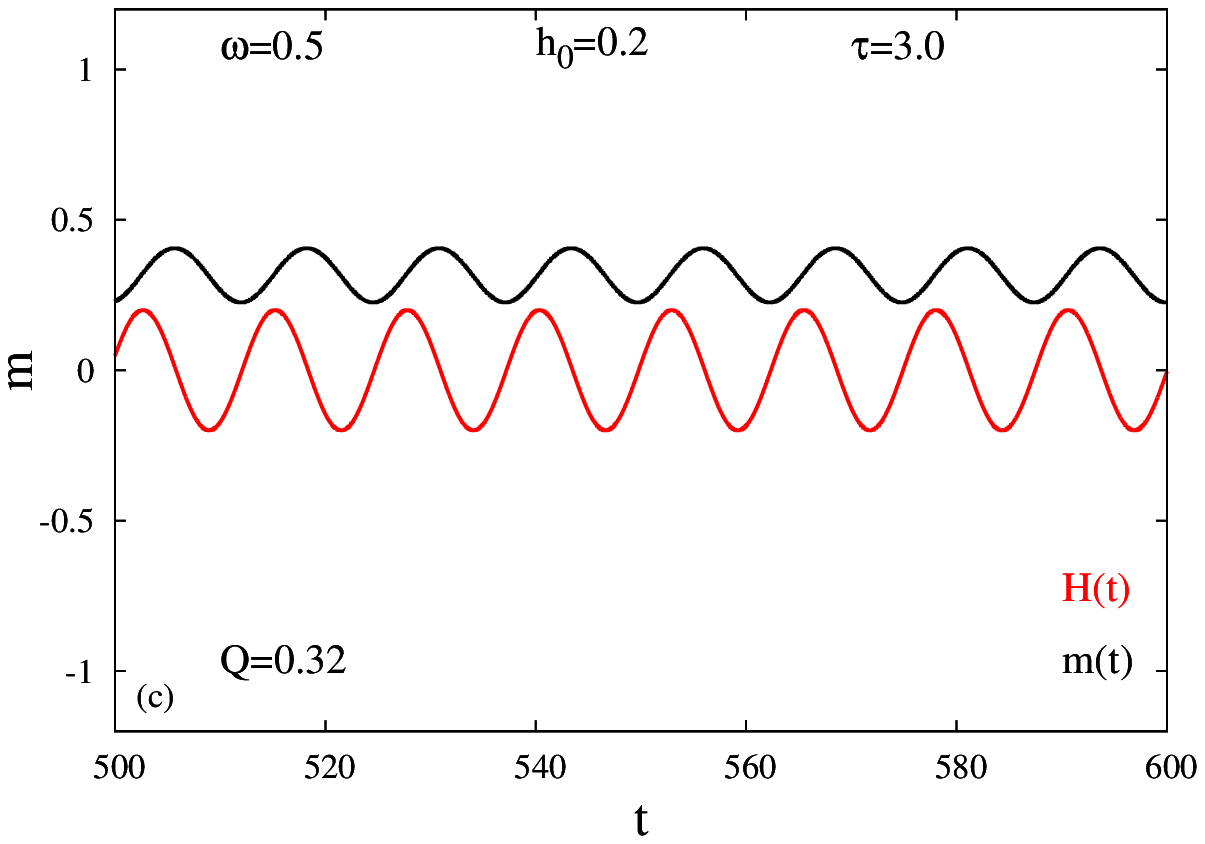, width=7.4cm}
\epsfig{file=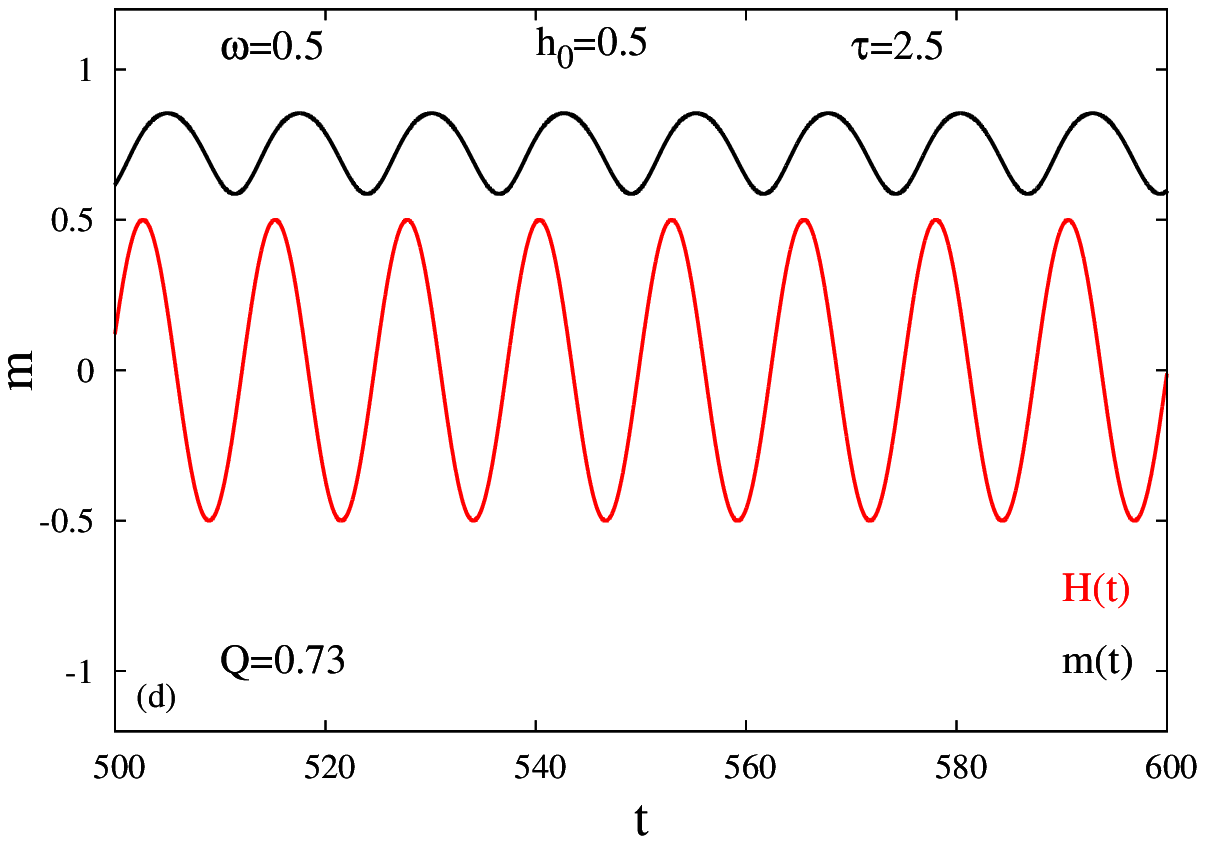, width=7.4cm}
\end{center}
\caption{Time series of the magnetization for the unmodulated field ($h_1=0.0$) 
with selected values of frequency, amplitude and temperature.} 
\label{sek1}\end{figure}

In order to construct the foundation of our discussion about the amplitude 
modulation, let us review unmodulated case briefly, i.e. $H_1=0.0$ in Eq. 
\re{denk2}. We can see several time series of the magnetization for different 
values of the Hamiltonian parameters and temperature in Fig. \ref{sek1}. On 
each 
figure the values of these parameters have been denoted. Time series are 
obtained by Runge-Kutta method which was used for the solution of Eq. 
\re{denk11}.  In Fig. \ref{sek1} (a) the system is in a dynamically disordered 
state, 
the value of the dynamical order parameter $Q=0.0$, magnetization oscillates 
around the zero value. When the value of the frequency increases, the system 
becomes unable to follow the magnetic field, then the system becomes ordered. 
This situation can be seen in Fig. \ref{sek1} (b). Also, we can see by 
comparing 
  Fig. \ref{sek1} (a) with (c) that, decreasing amplitude of the field  has similar 
effect. Decreasing amplitude means, decreasing energy supplied to the 
system by magnetic field, thus the spin-spin interaction overcome this effect. 
Lastly, decreasing temperature, due to the decreasing thermal fluctuations may 
result in ordered phase. This last observation can be seen by comparing Fig. 
\ref{sek1} (a) with (d). These are very well known results in the 
literature. 

\subsection{Modulated case with $h_0=0.0$}\label{results2}

\begin{figure}[h]\begin{center}
\epsfig{file=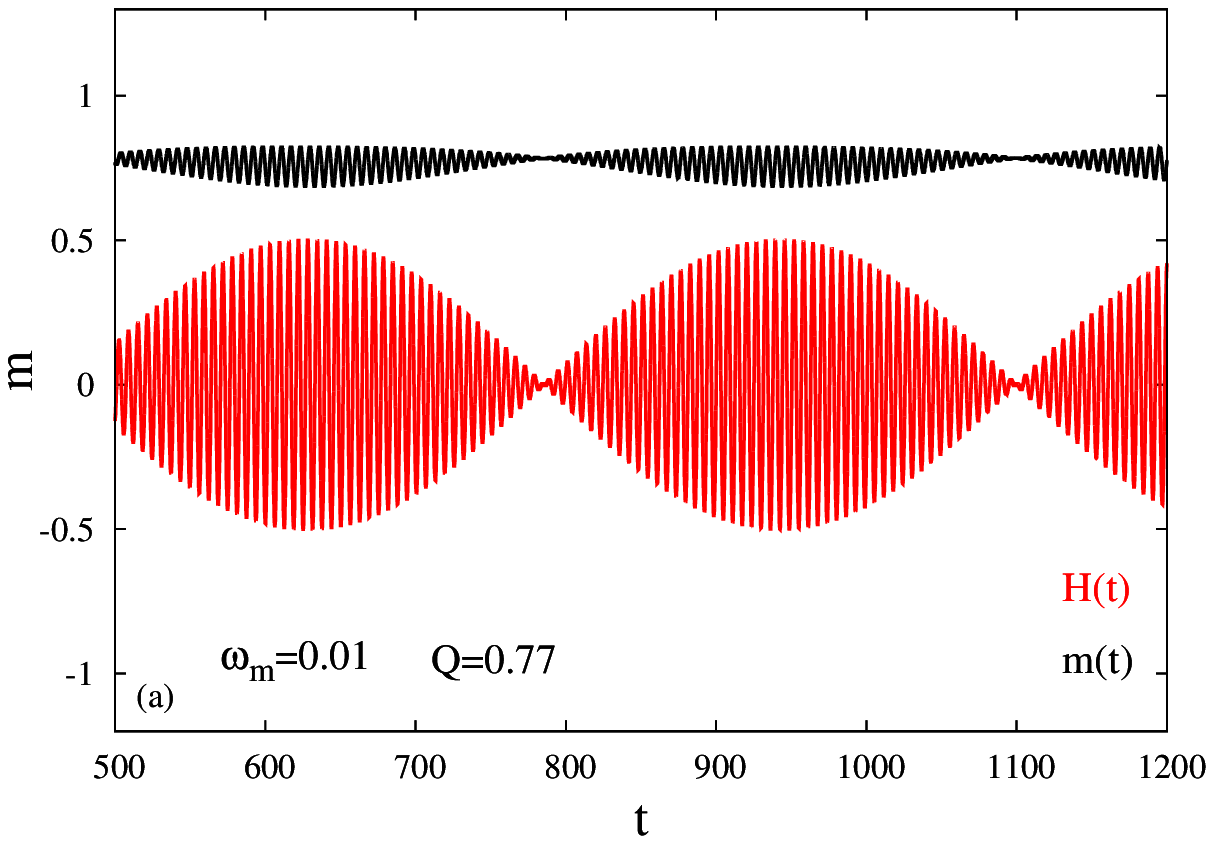, width=7.4cm}
\epsfig{file=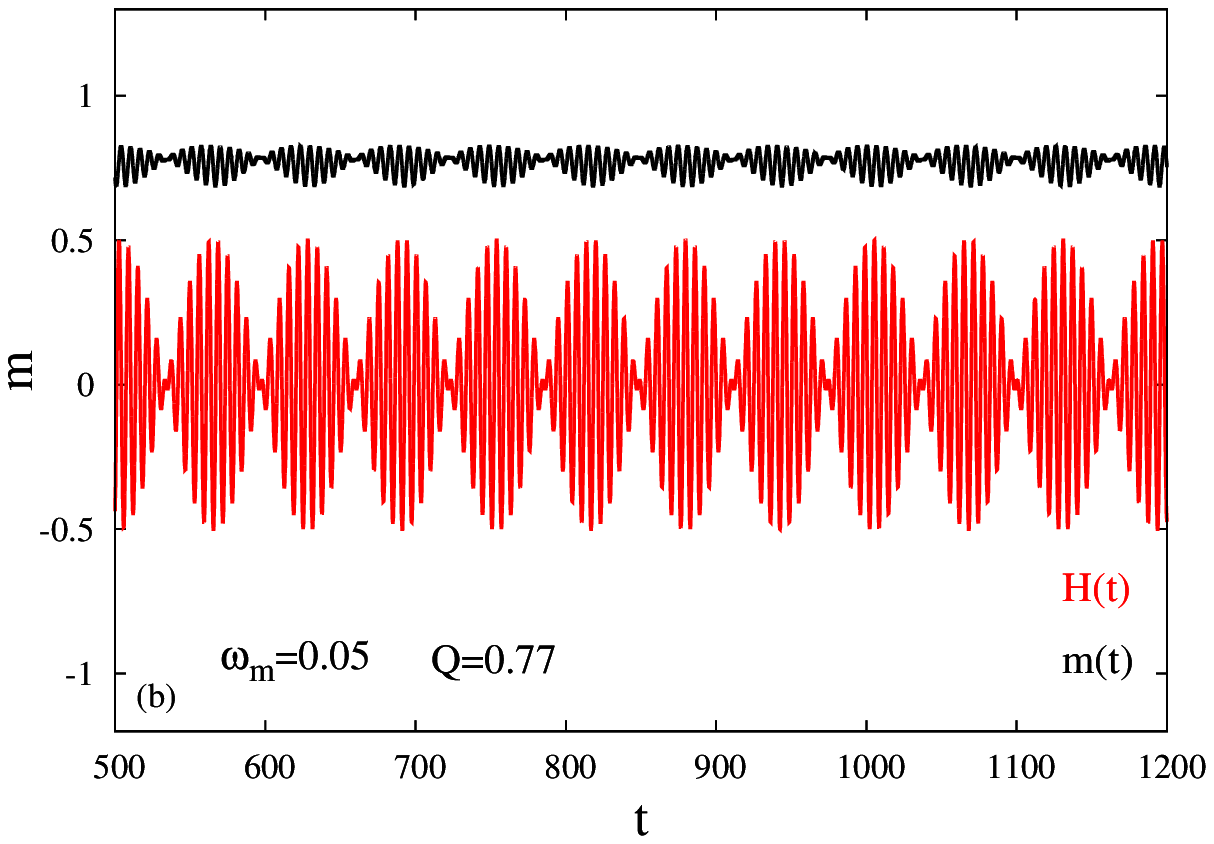, width=7.4cm}
\epsfig{file=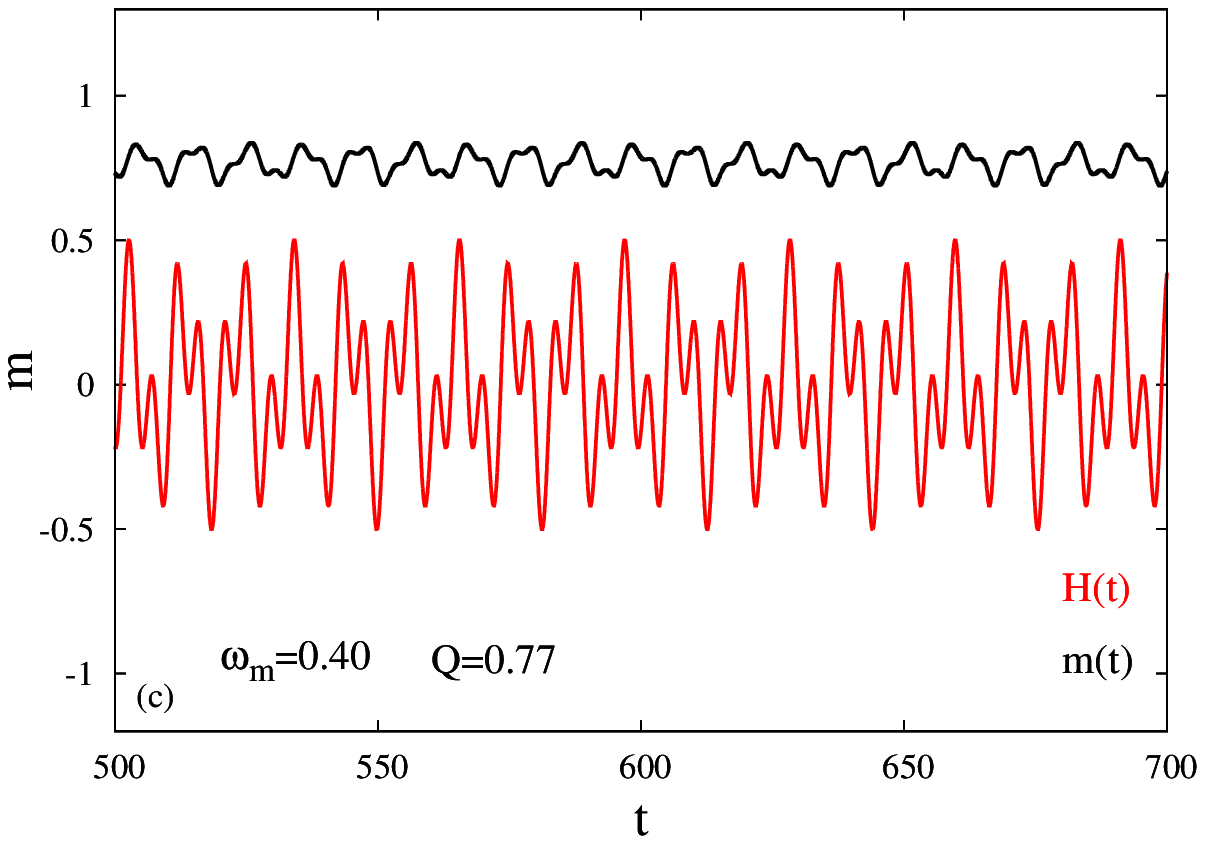, width=7.4cm}
\epsfig{file=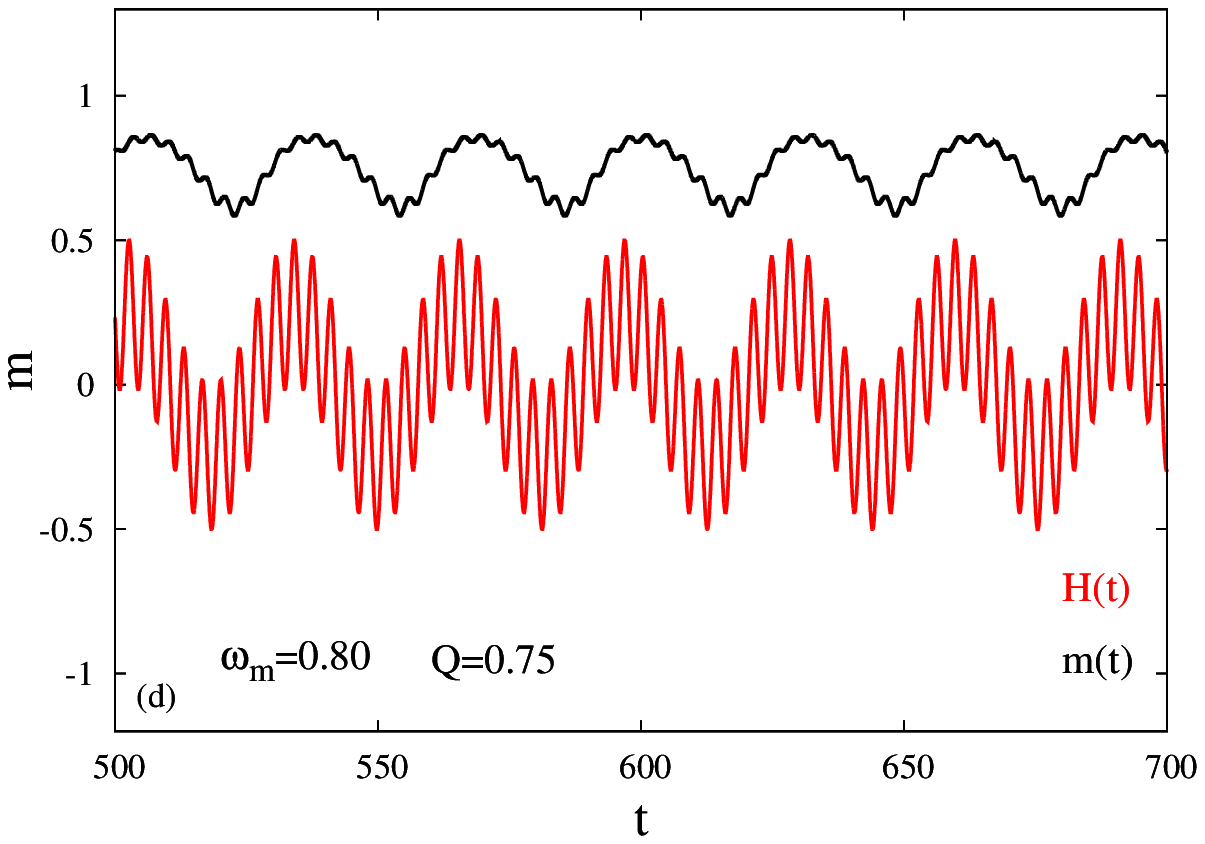, width=7.4cm}
\epsfig{file=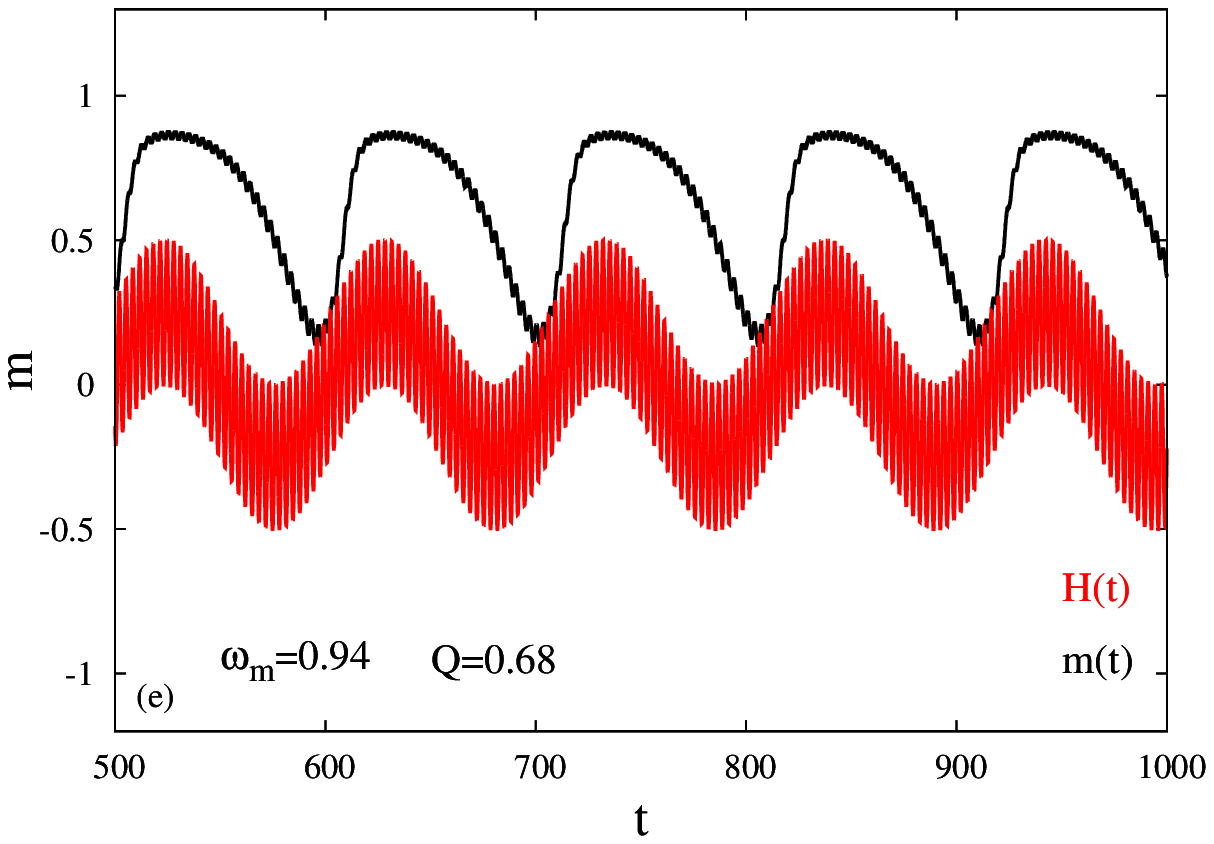, width=7.4cm}
\epsfig{file=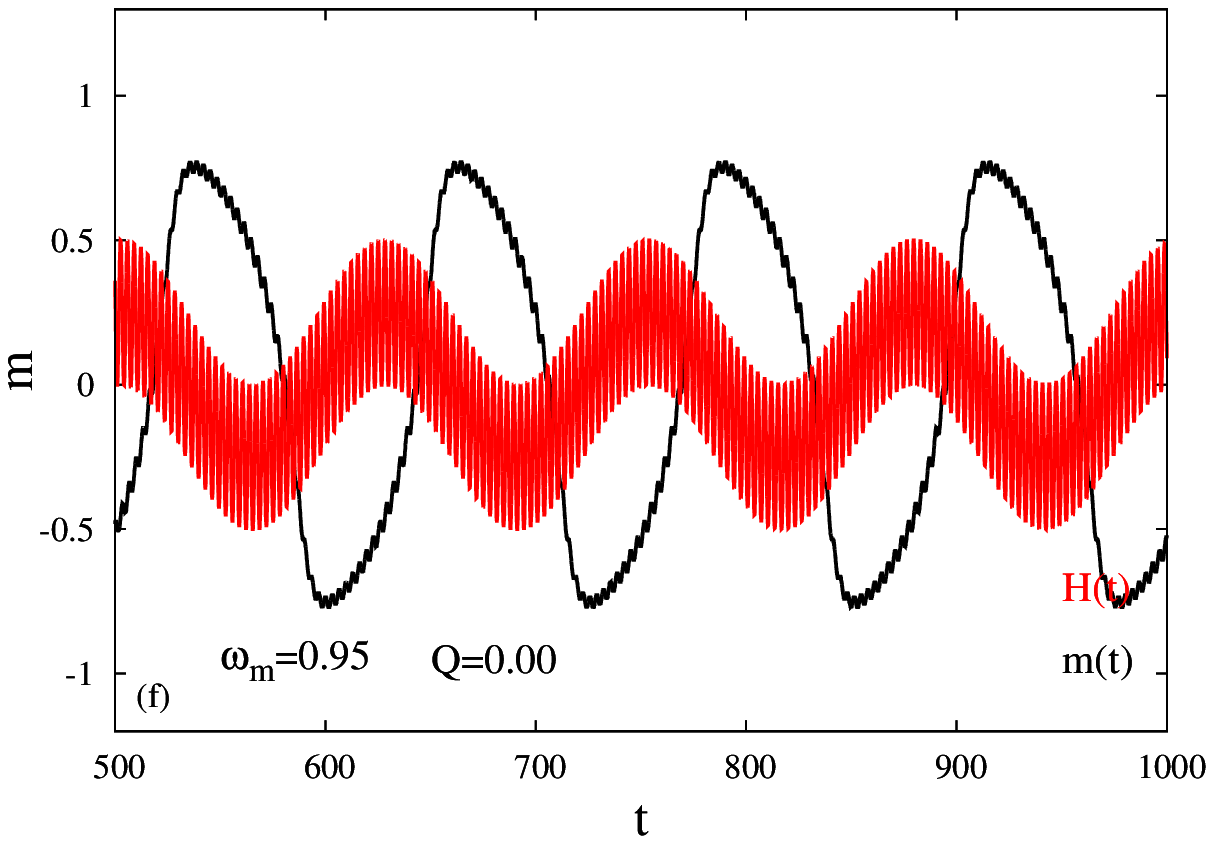, width=7.4cm}
\end{center}
\caption{Time series of the magnetization for the modulated field with selected 
values of frequency $\omega_m$. Other parameter values fixed as $h_0=0.00, 
h_1=0.50$, $\omega=1.00$ and $\tau=2.50$} 
\label{sek2}\end{figure}

Simpler case of modulation given by Eq. \re{denk2}, is the case with $H_0=0.0$, 
named as multiplier modulation. The (Fourier) spectrum of this field consists 
of 
frequencies $\omega \pm \omega_m , -\omega \pm \omega_m $. Since the aim is 
to determine the effect of the modulation, we have fixed the frequency of the 
unmodulated field as $\omega=1.0$ and $h_1=0.5$. Note that, due to the fact that field 
consist of product of two cosines, it will be enough to inspect the modulation 
frequency range $0.0<\omega_m<1.0$. The effect of the modulation can be seen in 
Fig. \ref{sek2}, where are time series of the field and magnetization for 
several values of the modulation frequency $\omega_m$ has been plotted. The value of the 
dynamical order parameter $Q$ is given in each case. As seen in \sek{sek2},  
changing the frequency of the modulation changes the behavior of the 
magnetization drastically. For lower values of the $\omega_m$, both of field 
and 
magnetization have wave packet like behavior (see \sek{sek2} (a) and (b)). 
System is in an ordered state with the value of dynamical order parameter 
$Q=0.77$. When the value of the modulation frequency rises, the view of both 
field and magnetization changes while the value of the order parameter is the 
same (see \sek{sek2} (c)). If the value of the  $\omega_m$ still increases, the 
value of the order parameter starts to decrease (compare \sek{sek2} (c) with 
(d)) and drop to zero (see \sek{sek2} (f)).The modulation could create 
dynamical 
phase transition for fixed values of amplitude and temperature.

For a closer look at the variation of the dynamical order parameter  with the 
modulation frequency we depict it for several values of temperature and 
amplitude, which can be seen in \sek{sek3}. At first sight symmetry about the 
$\omega_m=1.0$ takes attention. In general this symmetry is about the 
$\omega_m=\omega$ (remember that we have fixed the value of the modulation 
frequency as $\omega_m=1.0$).


As seen in \sek{sek3} abrupt change of value of dynamical order  parameter to 
the value of zero, leave place to smooth change when the temperature rises 
(compare curves related to the $\tau=1.5$ with $\tau=1.8$ in \sek{sek3} (a) or 
curves related to the $\tau=1.0$ with $\tau=1.5$ in \sek{sek3} (b)). Second 
effect of the rising temperature is to widen the plateau of $Q=0.0$.

\begin{figure}[h]\begin{center}
\epsfig{file=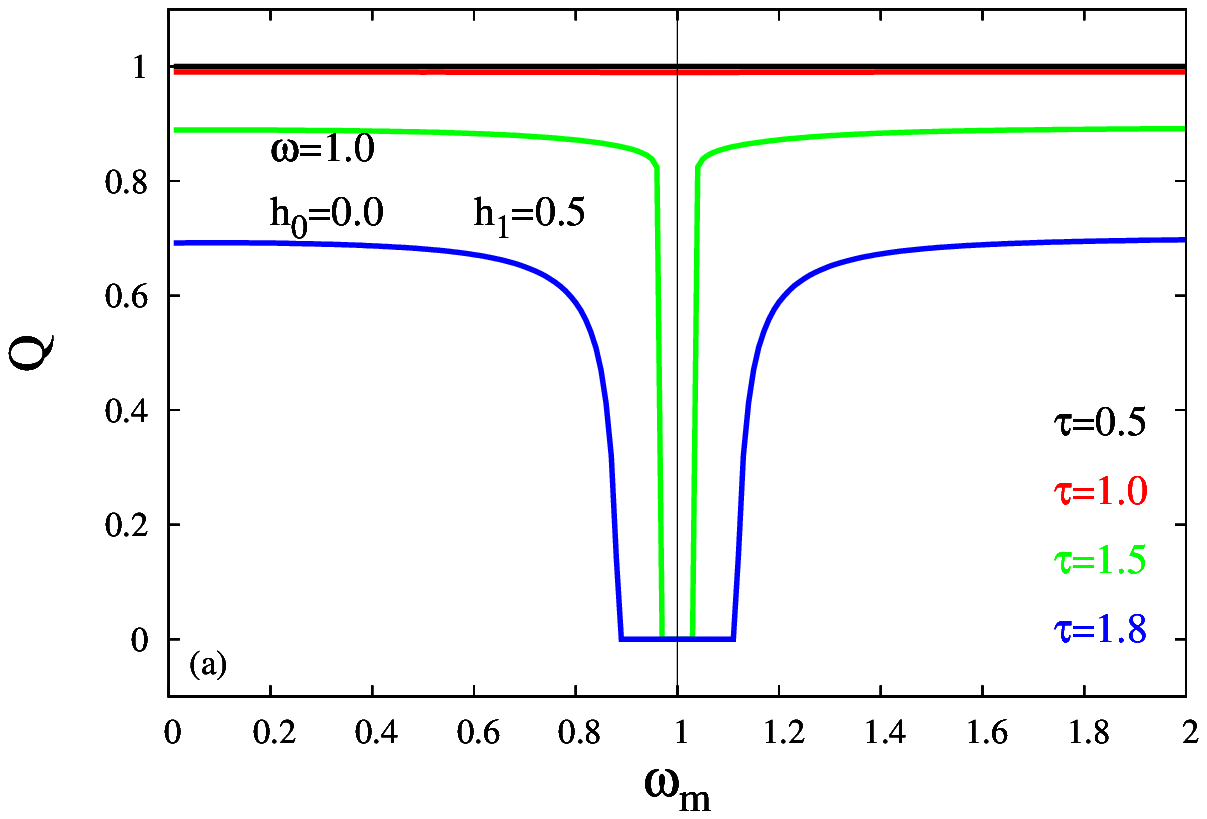, width=7.4cm}
\epsfig{file=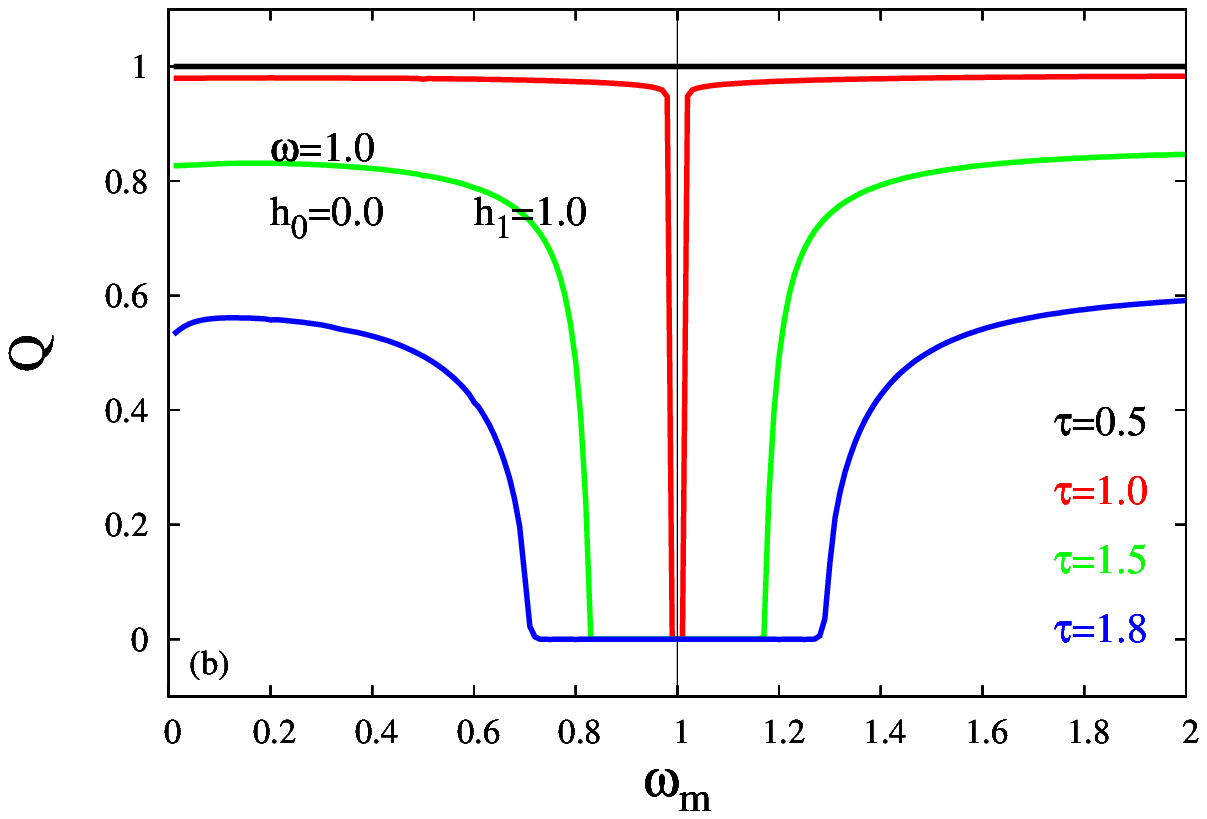, width=7.4cm}
\end{center}
\caption{Variation of the dynamical order parameter with the modulation 
frequency for the parameter values of $\omega=1.0$, $h_0=0.0$ and for chosen 
values of the temperature for amplitudes (a) $h_1=0.5$ and (b) $h_1=1.0$} 
\label{sek3}\end{figure}

\subsection{Modulated case with $h_0>0.0$}\label{results2}

This case includes fields that satisfy $H_0>0.0$. We can see from \re{denk2} that, the 
field term $h_0\cos\paran{\omega t}$ survives, regardless of the choice of 
$\omega_m$ and $h_1$.  

In order to determine the behavior of the magnetization with modulation 
frequency 
we depict the time series of the applied modulated field and magnetization for  
selected values of modulation frequency, as can be seen in \sek{sek4}. Other 
parameter values are fixed  
as $h_0=0.25, h_1=0.25, \omega=1.00$ and $\tau=2.50$. As seen in \sek{sek4}, 
rising modulation frequency drastically change the behavior of magnetization in 
time, as in the case of modulation with $h_0=0.0$. Again, for lower frequencies, 
wave packet like behavior takes attention. Rising frequency changes this 
behavior. In contrast to the case  $h_0=0.0$, modulation with $h_0>0.0$ could 
not create dynamically disordered phase (compare \sek{sek2} (e), (f) with 
\sek{sek4} (e), (f)). This is due to surviving term $h_0\cos\paran{\omega t}$ 
regardless of the value of modulation frequency. 

\begin{figure}[h]\begin{center}
\epsfig{file=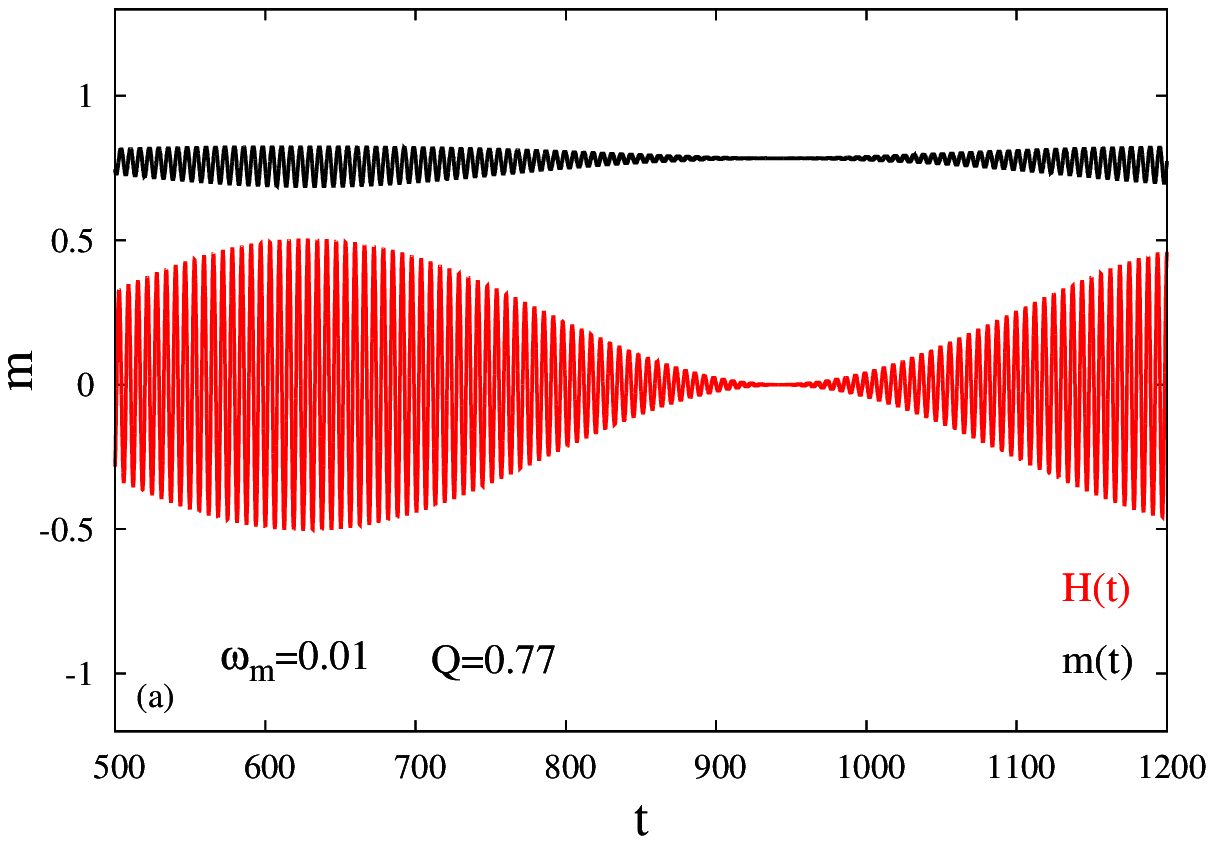, width=7.4cm}
\epsfig{file=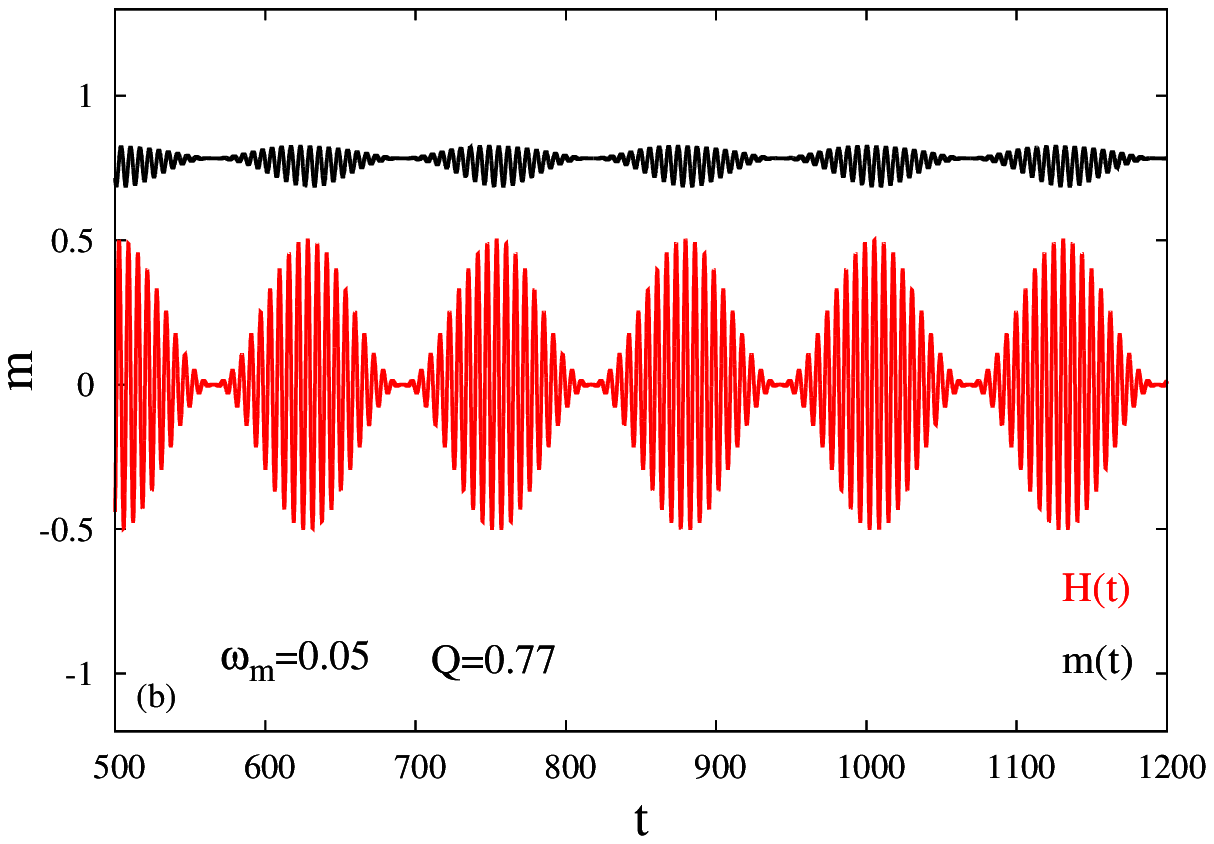, width=7.4cm}
\epsfig{file=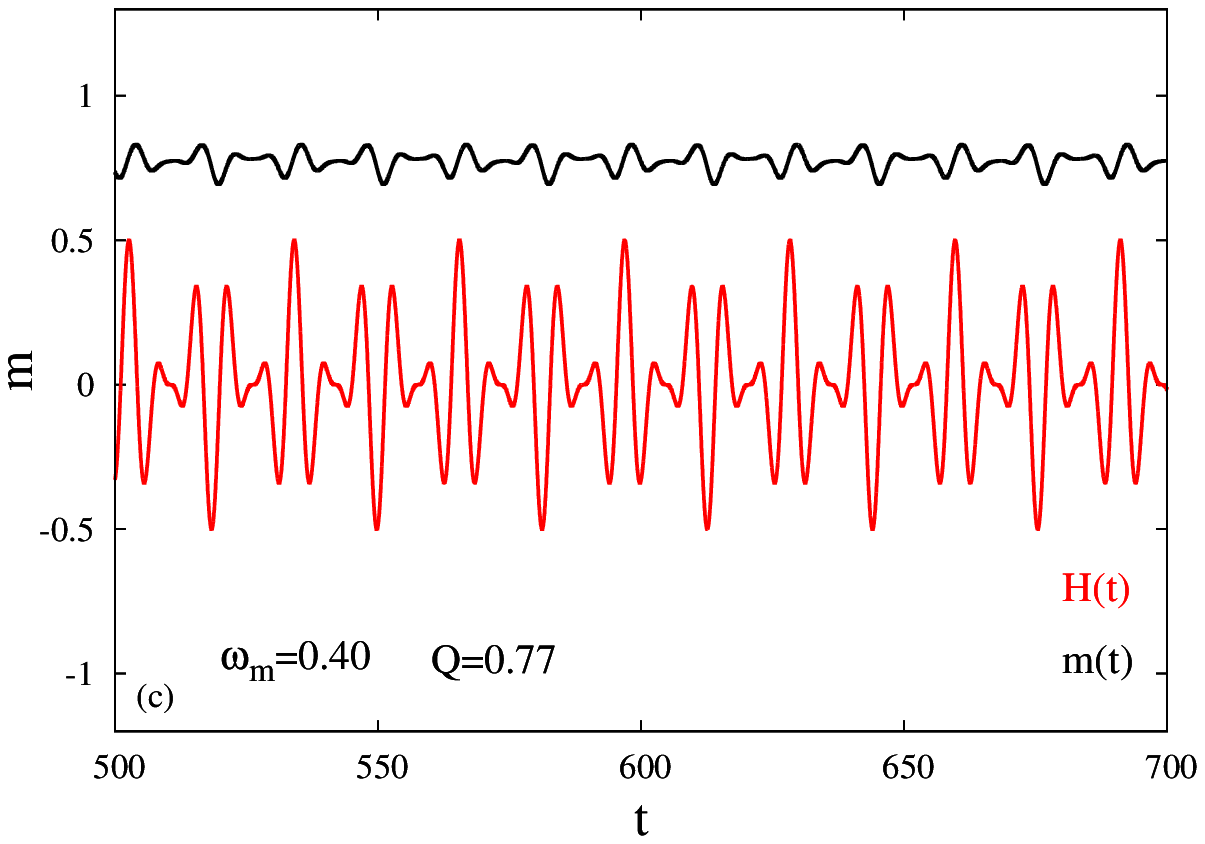, width=7.4cm}
\epsfig{file=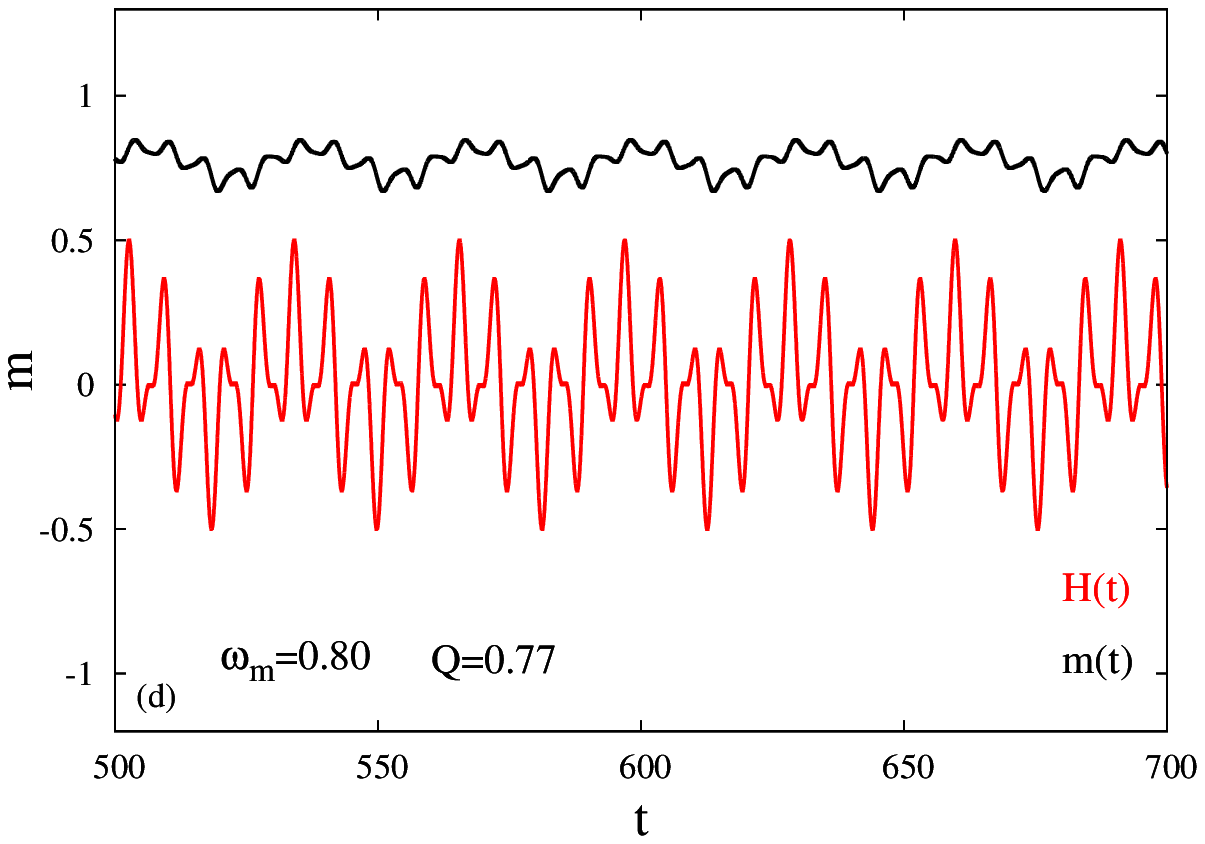, width=7.4cm}
\epsfig{file=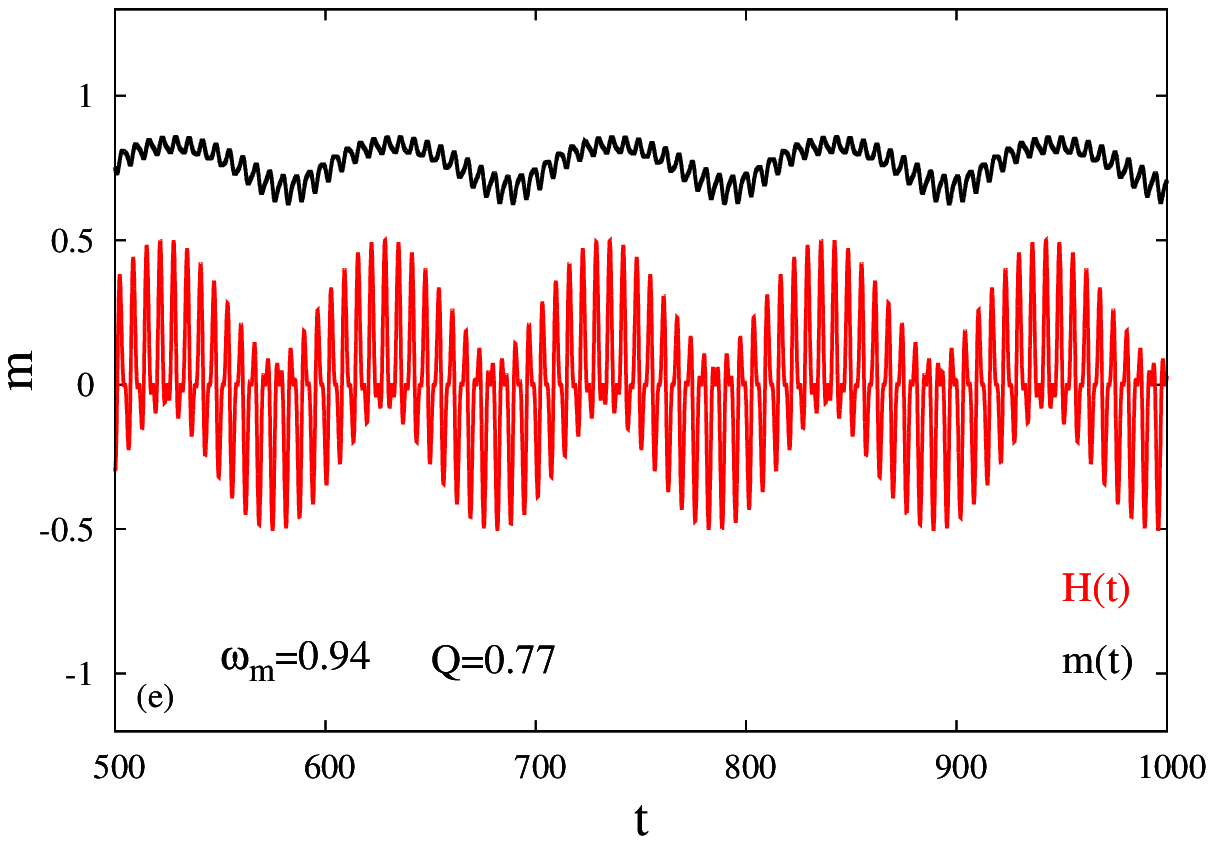, width=7.4cm}
\epsfig{file=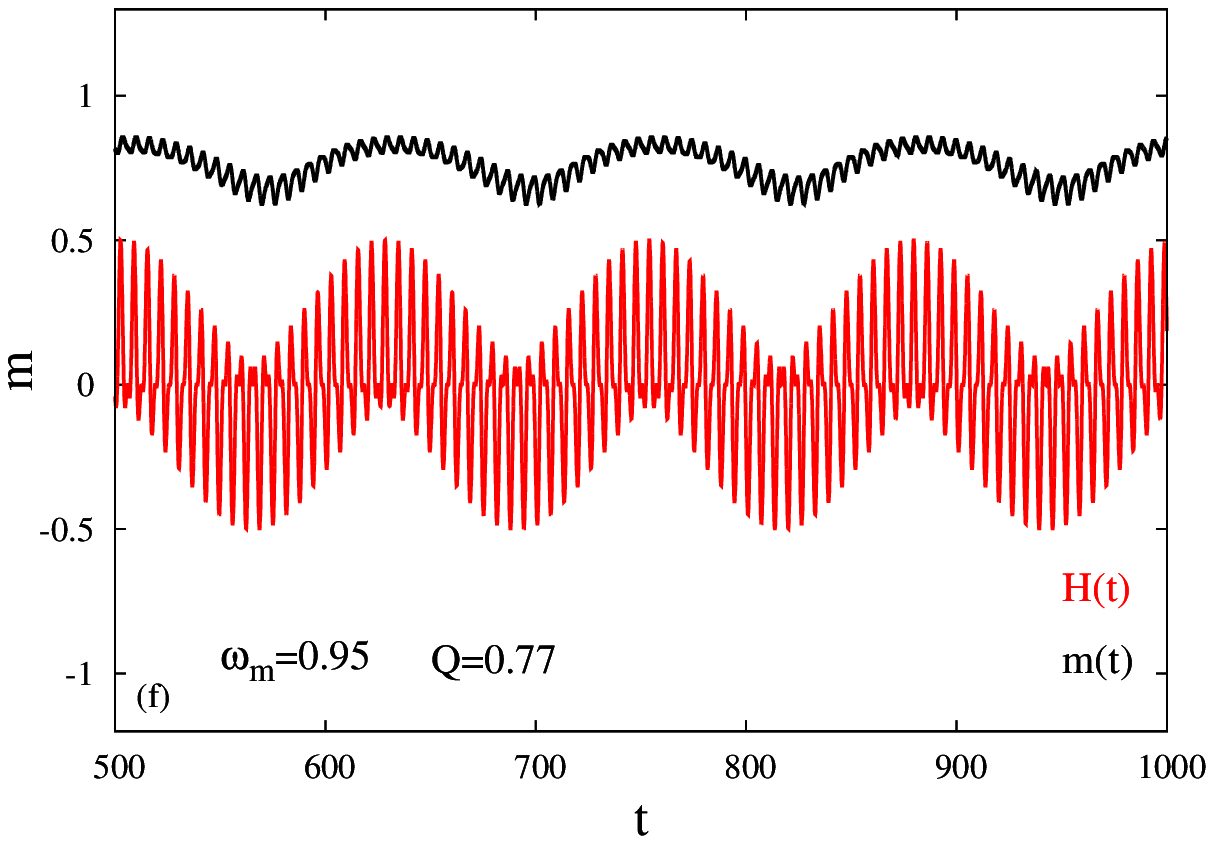, width=7.4cm}
\end{center}
\caption{Time series of the magnetization for the modulated field with selected 
values of frequency $\omega_m$. Other parameter values fixed as $h_0=0.25, 
h_1=0.25$, $\omega=1.00$ and $\tau=2.50$} 
\label{sek4}\end{figure}

\section{Conclusion}\label{conclusion}

The dynamical Ising model under the effect of the amplitude modulated time 
dependent periodic magnetic field has been solved
by using EFT with Glauber type of stochastic process. Since DPT in time 
dependent periodic magnetic field is well known, 
values of some Hamiltonian parameters fixed, such as frequency and temperature. 
Detailed investigation on
time series constructed with  several amplitude modulation case has been carried 
out. It has been 
shown that in the multiplier amplitude modulation case, rising modulation 
frequency could create DPT. The modulation frequency and 
frequency of the magnetic field has inverse effect on dynamical phases of the 
system. While rising modulation frequency could create 
dynamically disordered phase (as shown in this work), rising frequency drag the 
system to the ordered phase. This is because of
the fact that, since the frequency gets higher, magnetic system could not follow 
the driving field. Thus we can say that when the 
modulation  frequency gets higher, magnetic  system has the chance to follow 
the driving modulated magnetic field.

We hope that the results  obtained in this work may be beneficial 
form both theoretical and experimental point of view.

\section*{Acknowledgements}\label{ack}

The author would like to thank to \c{S}ebnem Se\c ckin U\u gurlu from Dokuz 
Eyl\" ul University for valuable discussions on frequency 
and amplitude modulation.

\end{document}